\RequirePackage{fix-cm}
\documentclass[twocolumn,epjc3]{svjour3} 
\usepackage[numbers]{natbib}

\pdfoutput=1 
\usepackage{graphicx}	
\usepackage{dcolumn}	
\usepackage{bm}			

\usepackage{color} 		
\usepackage[english]{babel}

\RequirePackage{graphicx}
\RequirePackage{color}
\RequirePackage[colorlinks,citecolor=blue,urlcolor=blue,linkcolor=blue]{hyperref}
\usepackage{multirow}
\usepackage[T1]{fontenc}
\usepackage[latin1]{inputenc}
\usepackage{amsmath,amssymb}
\usepackage{arydshln}
\usepackage{microtype}
\usepackage[switch]{lineno}
 \usepackage{orcidlink}

\begin{document}

\title{First upper limits on neutrino electromagnetic properties from the  \textsc{CONUS} experiment} 

\author{H.~Bonet\thanksref{addr1}, A.~Bonhomme\thanksref{addr1}\orcidlink{0000-0002-0218-2835}, C.~Buck\thanksref{addr1}\orcidlink{0000-0002-5751-5289}, K.~F\"ulber\thanksref{addr2}, J.~Hakenm\"uller\thanksref{a, addr1}\orcidlink{0000-0003-0470-3320}, J.~Hempfling\thanksref{addr1}\orcidlink{0000-0003-1191-0133}, G.~Heusser\thanksref{addr1}, T.~Hugle\thanksref{addr1}\orcidlink{0000-0001-9788-4014}, M.~Lindner\thanksref{addr1}\orcidlink{0000-0002-3704-6016},  W.~Maneschg\thanksref{addr1}\orcidlink{0000-0003-0320-7827}, T.~Rink\thanksref{addr1}\orcidlink{0000-0002-9293-1106}, H.~Strecker\thanksref{addr1}, R.~Wink\thanksref{addr2} (CONUS Collaboration)}

\thankstext{a}{e-mail: conus.eb@mpi-hd.mpg.de}

\institute{Max-Planck-Institut f\"ur Kernphysik, Saupfercheckweg 1, 69117 Heidelberg, Germany \label{addr1}
           \and
          PreussenElektra GmbH, Kernkraftwerk Brokdorf, Osterende, 25576 Brokdorf, Germany \label{addr2}
}

\date{Received: date / Accepted: date}
\maketitle

\abstract{
We report first constraints on neutrino electromagnetic properties from neutrino-electron scattering using data obtained from the \textsc{CONUS} germanium detectors, i.e.~an upper limit on the effective neutrino magnetic moment and an upper limit on the effective neutrino millicharge. The electron antineutrinos are emitted from the 3.9\,GW$_\mathrm{th}$ reactor core of the Brokdorf nuclear power plant in Germany. The \textsc{CONUS} low background detectors are positioned at 17.1\,m distance from the reactor core center. The analyzed data set includes 689.1\,kg$\cdot$d collected during reactor ON periods and 131.0\,kg$\cdot$d collected during reactor OFF periods in the energy range of 2 to 8\,keV$_{ee}$\footnote{The subscript "$ee$" refers to ionization energy, which is the energy directly collected by the Ge spectrometers.}. With the current statistics, we are able to determine an upper limit on the effective neutrino magnetic moment of $\mu_\nu < 7.5\cdot10^{-11}\,\mu_B$ at 90\% confidence level. No neutrino signal in this channels as well as in the CE$\nu$NS channel has been observed at a nuclear power plant so far. From this first magnetic moment limit we can derive an upper bound on the neutrino millicharge of $\vert q_{\nu}\vert < 3.3\cdot10^{-12}\,e_0$.

}


\section{Introduction}\label{sec:intro}
In the standard model (SM) neutrinos are massless and do not possess electric or magnetic dipole moments. However, today we know from many experiments that neutrinos have mass. This observation demands for physics beyond the standard model (BSM). In minimal extensions of the SM with three right-handed neutrinos the magnetic moment $\mu_\nu$ is predicted to be non-zero with values of less than $10^{-14}-10^{-15}\mu_B$~\cite{PhysRevLett.95.151802, Fujikawa:1980yx, PhysRevD.39.3378, Giunti:2014ixa}, with the Bohr magneton $\mu_B$ as conventional unit.  
This model-independent limit takes into account the natural contribution to the neutrino mass, i.e.\ $\delta m_{\nu} \lesssim m_{\nu}$.
Much weaker constraints to values of $10^{-9}\mu_B$ are obtained for theoretical models assuming Majorana neutrinos in a comparable way~\cite{shrock1982electromagnetic, Bell:2006wi}. An experimental discovery of a neutrino magnetic moment (NMM) with values larger than $> 10^{-15}\mu_B$ would therefore hint towards neutrinos being Majorana fermions. 
Moreover, electromagnetic interactions of neutrinos are highly relevant in astrophysical environments \cite{cisneros1971effect}, where neutrinos are exposed to magnetic fields on large scales.

The electromagnetic differential cross-section for the neutrino-electron scattering under the assumption of a NMM is

\begin{equation}
	\left(\frac{d\sigma}{dT}\right)_\mathrm{em} = \pi \frac{\alpha^{2}}{m_{e}^{2}} \left(\frac{1}{T} - \frac{1}{E_{\nu}}\right) \left(\frac{\mu_{\nu}}{\mu_{B}}\right)^{2}\, ,
	\label{nmmcrosssection}
\end{equation}

with the neutrino energy $E_{\nu}$ and the electron recoil energy $T$. The electron mass is denoted by $m_{e}$ and $\alpha$ refers to the fine structure constant.

Experimental limits on $\mu_\nu$ have been determined from different neutrino sources. Beyond supernova neutrino constraints \cite{barbieri1988limit}, the best limit so far is provided by searches for elastic scattering of solar neutrinos off electrons in the \textsc{BOREXINO} liquid scintillator detector~\cite{Borexino:2017fbd}. Here an upper limit on the effective NMM of $\mu_\nu < 2.8\cdot 10^{-11}\mu_B$ was determined at 90\% confidence level (C.L.). Similar or less stringent constraints were derived from several commercial reactor antineutrino experiments~\cite{Beda:2013mta, TEXONO:2006xds, Derbin:1993wy}. Neutrino-electron scattering was also studied in accelerator experiments. However, in general, the expected sensitivity is higher for reactor-based investigations than for pion-decay-at-rest sources as used in the \textsc{COHERENT} experiment~\cite{COHERENT:2017ipa}, since lower electron recoils can be explored with reactor neutrinos at a very high flux~\cite{Miranda:2019wdy}. 
Moreover, with the detection of CE$\nu$NS at reactor site, which is still pending, a significant improvement in the limit provided by reactor experiments, is expected from the analysis of the low-energetic recoil spectrum \cite{billard2018prospects}. 
For muon neutrinos not available at reactor-site, accelerator experiments provide the best bounds~\cite{Allen:1992qe, Auerbach:2001wg}. It should be noted that the constraints obtained in those experiments refer to an effective NMM that differs from e.g.~solar or reactor neutrinos~\cite{Giunti:2014ixa}. Thus, there is a complementarity between the different experimental approaches, in particular as regards the role of Dirac and Majorana CP phases~\cite{Canas:2015yoa}. Recently, the \textsc{XENON} Collaboration reported an excess of events at low energies between 2 and 3\,keV in their dual-phase noble gas detector designed for dark matter searches~\cite{Aprile:2020tmw}. Among other explanations, this excess can be interpreted as a signal from solar neutrinos with an enhanced NMM, just slightly below the current best limit from \textsc{BOREXINO}. However, there are already tensions with the constraints from astrophysical limits. Those are about one order of magnitude stronger, but model-dependent~\cite{Arceo-Diaz:2015pva, Viaux:2013lha, Kuznetsov:2009zm}. 

Since neutrinos have mass, it is not obvious that their electric charge is zero as assumed in the SM \cite{giunti:2007}. 
Instead, neutrinos might be electrically millicharged particles. An estimate on the scale of this millicharge for the electron antineutrino (NMC) can be made based on the determined limit on the effective NMM and the electron recoil energy threshold $T$ as applied in~\cite{Studenikin:2013my}. The corresponding limit on $q_{\nu}$ can be estimated as follows~\cite{Studenikin:2013my}: 

\begin{equation}
	q_{\nu}^2<\frac{T}{2m_e}\left(\frac{\mu_{\nu}}{\mu_B}\right)^2 e_0\, ,
	\label{eq:millicharge}
\end{equation}

with $m_e$ being the electron mass and $e_0$ the elementary charge. 

A commercial nuclear reactor is a promising source to study electromagnetic properties of neutrinos, mainly due to its high flux of more than $10^{20}$~neutrinos per second at GW thermal power, the favorable neutrino energy range up to about 10\,MeV as well as the well-defined spectrum and source location. Although experiments close to nuclear reactors are in principle capable to measure both - the neutrino-electron and the neutrino-nucleus scattering channel, the technologies of the running experiments are mainly sensitive to the former interaction whose signature extends towards higher energies. As a potential signal increases as $T^{-1}$, a detector technology with low detector threshold and good background control is required. Therefore, the currently planned and running reactor neutrino projects searching for coherent elastic neutrino-nucleus scattering (CE$\nu$NS), like the \textsc{CONUS} experiment~\cite{Bonet:2020awv}, provide an ideal environment for such type of studies.


\section{The \textsc{CONUS} experiment}\label{sec:experiment}

The \textsc{CONUS} experiment is set up at a distance of (17.1$\pm$0.1)\,m to the reactor core of the nuclear power plant in Brokdorf (KBR), Germany. The close distance to the reactor core as well as the maximum thermal power of 3.9\,GW$_\mathrm{th}$ ensure a high antineutrino flux. The setup consists of a compact shield enclosing four high-purity point contact germanium (Ge) spectrometers (C1, C2, C3 and C4) with a total active mass of (3.73$\pm$0.02)\,kg \cite{Bonet:2020ntx}. 
The data collecting periods are split into reactor ON and OFF times as well as commissioning and optimization phases. 
The physics data collection started in April 2018. There is one regular outage lasting three to four weeks per year.

The Ge spectrometers exhibit very low electronic noise, which becomes evident from the resolution of artificially generated pulse signals passing through the electronics chain of 60 to 80\,eV$_{ee}$ full-width at half-maximum. This translates into a detection efficiency close to 100\% down to 200\,eV$_{ee}$. All four detectors have a similar active mass within [0.91,0.95]\,kg, in which all energy depositions are reconstructed at the correct energy.

The most relevant background contributions of the experiment are: natural radioactivity and airborne radon, contaminations in the detector and shield material including cosmic activation, muon-induced background and potentially reactor-induced background. 
All of them are suppressed by a careful selection of materials, an elaborated onion-like shield design and an active anticoincidence muon veto. An overview of the \textsc{CONUS} shield is given in \cite{bonet2021bkg}. An overall background level of 5-7\,d$^{-1}$kg$^{-1}$keV$^{-1}_{ee}$ is achieved in the energy region [2,8]\,keV$_{ee}$, which is about a factor of two smaller than the one of the GEMMA experiment.
The remaining measured background inside the shield is decomposed and described by a Monte Carlo (MC) simulation based background model \cite{bonet2021bkg}. An overview of all background contributions is provided in Table \ref{tab:detector_bkgoverview}. Within [2,8]\,keV$_{ee}$ it is dominated by recoils of muon-induced neutrons as well as signatures from decays of $^{210}$Pb and its progenies within the cryostat and the innermost shield layer. The spectral shape of the background created by the $^{210}$Pb decays in the vacancy of the diode is strongly influenced by surface effects in the charge collection. Especially the thickness of the so-called passivation layer on one side of the diodes is not well known, which leads to a shape uncertainty on the background model. This needs to be taken into account in the analysis. 

Next to the overall background description, a detailed understanding of any potential time-dependent background, especially if reactor-correlated, is highly relevant for the NMM search, since it could be misidentified as a signal.
The fluence rate of fission neutrons as well as the $\gamma$-ray background at the experimental site in room A-408 of the KBR were evaluated before the setup of the experiment in a dedicated measurement campaign \cite{hakenmuller2019neutron}. MC simulations demonstrate the reactor-correlated background to be negligible inside the detector chamber. The validation of the MC simulations is discussed in \cite{hakenmuller2019neutron} as well. 
Reactor-independent changes in the background can occur due to the decay of cosmogenic-induced lines and airborne radon. 
The cosmogenic-induced lines are created at Earth's surface in germanium and copper by the hadronic component of the cosmic rays. They can also be produced in neutron capture by mostly thermal neutrons. In the \textsc{CONUS} data, lines emitted by $^{68}$Ge, $^{71}$Ge, $^{65}$Zn, $^{68}$Ga and $^{57}$Co are observed in the measured energy range. For all other relevant isotopes, e.g. $^{3}$H, the activation is estimated from the production rates in literature and the tracked exposure of the materials. This component induces less than 6\% of the total background below 8\,keV$_{ee}$ and the changes over time were shown to be negligible for the analysis at hand. However, due to the large uncertainties in the production rates the uncertainty on the spectral shape can be significant and has therefore been included as systematic shape uncertainty in the background model as well.    
Airborne radon is kept away from the detectors by flushing the detector chamber with air from breathing air bottles, that have been stored at least two weeks for the radon to decay. During the first physics run, the background suppression capability of the flushing was not utilized to its full potential. Thus, a minor radon contribution is included in the background model spectra. It is assumed to be stable over time. During the outage of RUN-2, the flushing was improved in order to keep the detector chamber completely radon free.

\begin{table}[ht]
    \centering
    \begin{tabular}{l l   }\hline \hline
     background source&contribution   \\ \hline
     muon-induced inside shield   & 32-47\%                      \\
     (applied muon veto)          &                    \\
     $^{210}$Pb (inside cryostat and shield)    & 31-58\%                      \\
     muon-induced  neutrons in concrete     & 4-6\%                      \\
     cosmogenic (decay over time) & 3-6\%                      \\
     airborne radon (run-dep.)    & 0.3-13\%                      \\ \hdashline
     overall background rate &  5-7\,d$^{-1}$kg$^{-1}$keV$^{-1}_{ee}$ \\
     GEMMA experiment \cite{beda2007first} &  $\sim$15\,d$^{-1}$kg$^{-1}$keV$^{-1}_{ee}$              \\
     \hline\hline
    \end{tabular}
    \caption{Overview on the different sources contributing to the background in the energy range between 2 and 8\,keV$_{ee}$ and comparison to the GEMMA experiment. A range is given to cover the different detectors and runs. \label{tab:detector_bkgoverview}}
\end{table}

\section{Data analysis}\label{sec:methods}

The NMM is investigated by the electromagnetic interactions of reactor antineutrinos with electrons in the target material (for the cross section see formula~\eqref{nmmcrosssection}). Any electroweak contributions are subdominant. 

At the \textsc{CONUS} experimental site a neutrino flux of 2.3$\cdot$10$^{13}$cm$^{-2}$s$^{-1}$ \cite{Bonet:2020awv} is expected. About 7.2\,$\bar{\nu}_{e}$ \cite{beda2007first} are assumed to be emitted per fission from the $\beta$-decays of $^{235}$U, $^{238}$U, $^{239}$Pu and $^{241}$Pu.
To calculate the signal expectation, the shape of the reactor antineutrino emission spectrum is described by the Huber-M\"uller \cite{Huber:2011wv, Mueller:2011nm} model completed by the ab-initio calculations of Kopeikin \cite{Kopeikin:2003gu} for low neutrino energies. 
Further, we include the observed spectral distortion (the reactor "bump") measured by the Daya Bay experiment with their provided correction factors \cite{DayaBay:2016ssb}.
The models are adapted with the help of the run-specific fission fractions provided by KBR to properly take into account the time-dependent contributions of the main fission isotopes. Atomic electron binding effects are taken into account for the electrons involved in the scattering process. The procedure described in Ref. \cite{beda2007first} is applied with the value for the binding energies in Ge taken from LLNL Evaluated Atomic Data Library (EADL) \cite{perkins1991tables}. The spectral shape is approximately proportional to the inverse of the electron recoil energy, which implies that the energy bins at low energies have a larger impact on the analysis result. Finally, the signal expectancy is folded with the energy resolution of the detectors.

\begin{figure}
    \centering
    \includegraphics[width=0.47\textwidth]{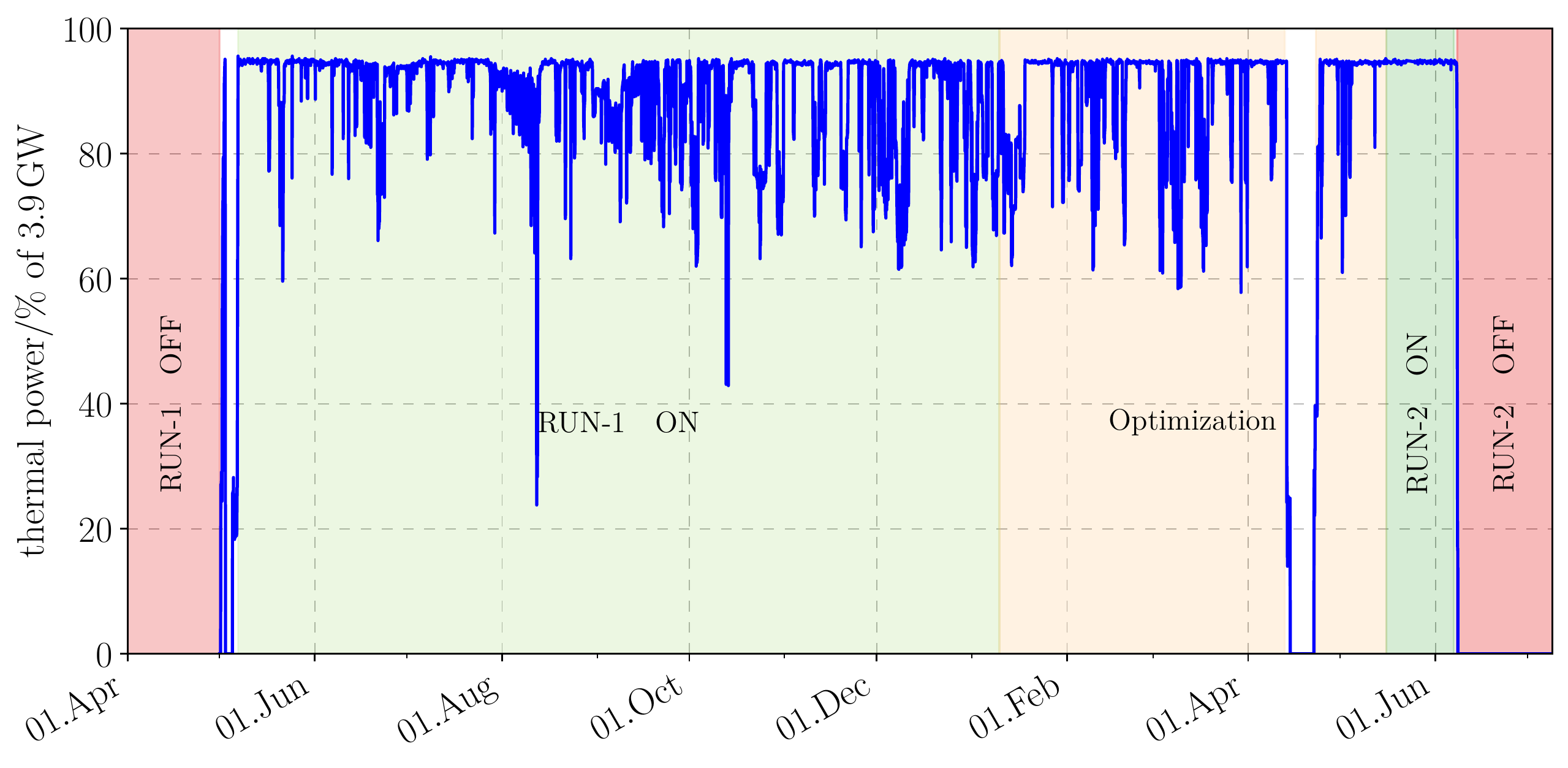}
    \caption{Relative reactor thermal power during RUN-1 and RUN-2 of the CONUS experiment in 2018 to 2019. The maximum absolute thermal power of KBR amounts to 3.9\,GW$_\mathrm{th}$. The outages are marked in red. The optimization period in yellow is not included in the data sets.}
    \label{fig:nmm_signalexpectation2}
\end{figure}

The selected data sets encompass RUN-1 and RUN-2 of the experiment. RUN-1 starts with the regular outage in April 2018 and ends in October 2018 with the beginning of an optimization phase including major changes in the settings of the data acquisition system (DAQ). RUN-2 includes reactor ON time from May 2019 and the following regular outage in June. The ending of this run is marked by a leakage test of the reactor safety vessel, which had an impact on the detector performance and background level. 
The thermal power output of the nuclear power plant during these periods is depicted in Figure \ref{fig:nmm_signalexpectation2}. The detector C4 is excluded from the analysis due to an artifact observed during RUN-1. 
The previously described improvements in the air flushing routine during the outage of RUN-2 demand an adaption of the radon-induced component in the background model. However, only for C3 a separate model is required for the reactor OFF and ON time because only this detector shows a significant impact of radon in the detector chamber on the data.    

To avoid any impact of unstable electronics noise on the data sets, the conservative limit on the lower end of the region of interest (ROI) is set at 2\,keV$_{ee}$ above the L shell X-ray lines of $^{68}$Ge, $^{68}$Ga and $^{75}$Zn \cite{bonet2021bkg}. At higher energies the ROI ends at 8\,keV$_{ee}$ below the K shell X-ray lines of these isotopes. The energy ranges of the individual detectors for the NMM and NMC analyses are listed in Table \ref{tab:detector_lifetimes}. A bin width of 50\,eV$_{ee}$ was selected for the analysis.
The complete exclusion of the electronics noise region enables us to loosen the noise-temperature correlation cut (NTC cut), that strongly impacted the exposure of the CE${\nu}$NS data set \cite{Bonet:2020awv}. This results in a total exposure of 689.1\,kg$\cdot$d for reactor ON and of 131.0\,kg$\cdot$d for reactor OFF as shown in Table \ref{tab:detector_lifetimes}. Nevertheless, the time difference distribution cut (TDD cut) is applied to the data sets to remove artifacts created by the DAQ \cite{Bonet:2020awv}.

\begin{table}[ht]
    \centering
    \begin{tabular}{c c c c c c }\hline \hline
     detector & run & ON [kg$\cdot$d] & OFF [kg$\cdot$d] & ROI [keV$_{ee}$]\\ \hline
     C1 & R1 & 215.4 & 29.6 & 2.013 - 7.968\\
     C2 & R1 & 184.6 & 32.2 & 2.006 - 7.990\\
     C3 & R1 & 248.5 & 31.7 & 2.035 - 7.989\\  \hline
     C1 & R2 &  19.8 & 18.5 & 2.010 - 7.955\\
     C3 & R2 &  20.8 & 19.0 & 2.007 - 7.990 \\ \hline
     all & all & 689.1    &  131.0    &  2 - 8 \\
     \hline\hline
    \end{tabular}
    \caption{Exposures and energy ranges of the performed neutrino magnetic moment investigation based on \textsc{CONUS} RUN-1 and RUN-2 data.\label{tab:detector_lifetimes}}
\end{table}

The treatment of the uncertainties is similar to the CE$\nu$NS analysis \cite{Bonet:2020ntx} as well as other BSM analyses from \textsc{CONUS}~\cite{bonet2021novel}. For the detectors and the DAQ system, systematic uncertainties on the active volume ($\sim$3\%), the energy scale ($\sim$10\,eV$_{ee}$) and the muon veto dead time ($\sim$0.1\%) are taken into account. 
Reactor-specific uncertainties are composed of uncertainties on the distance to the reactor core ($<$1\%), the reactor thermal power (2.3\%) and the uncertainty on the fission fractions provided individually for each run from the nuclear power plant. The uncertainty on the shape of the reactor-spectrum is assumed to have a minor impact and is therefore not taken into account. The previously described shape uncertainties of the background model amount to $\sim$5\% per bin at maximum. 

We carried out a binned likelihood fit, where all detectors and runs are simultaneously combined. The upper limit at 90\% C.L.~is extracted via a likelihood ratio test. The corresponding distribution of the test statistics is derived by a toy MC. All systematic uncertainties are included into the likelihood via Gaussian pull terms. 
The treatment of the shape uncertainty of the background model corresponds to an extension of the normalization parameter ($b_{0}$ parameter) with two additional parameters for a second order polynomial ($b_{1}\cdot E+b_{2}\cdot E^{2}$). While the normalization parameter is free in the fit, the other two parameters are restricted by Gaussian pull terms. In this way, the deviation averaged over all bins from the fit to the MC background model is restricted.


\section{Results and discussion}

\begin{figure}
    \centering
    \includegraphics[width=0.5\textwidth]{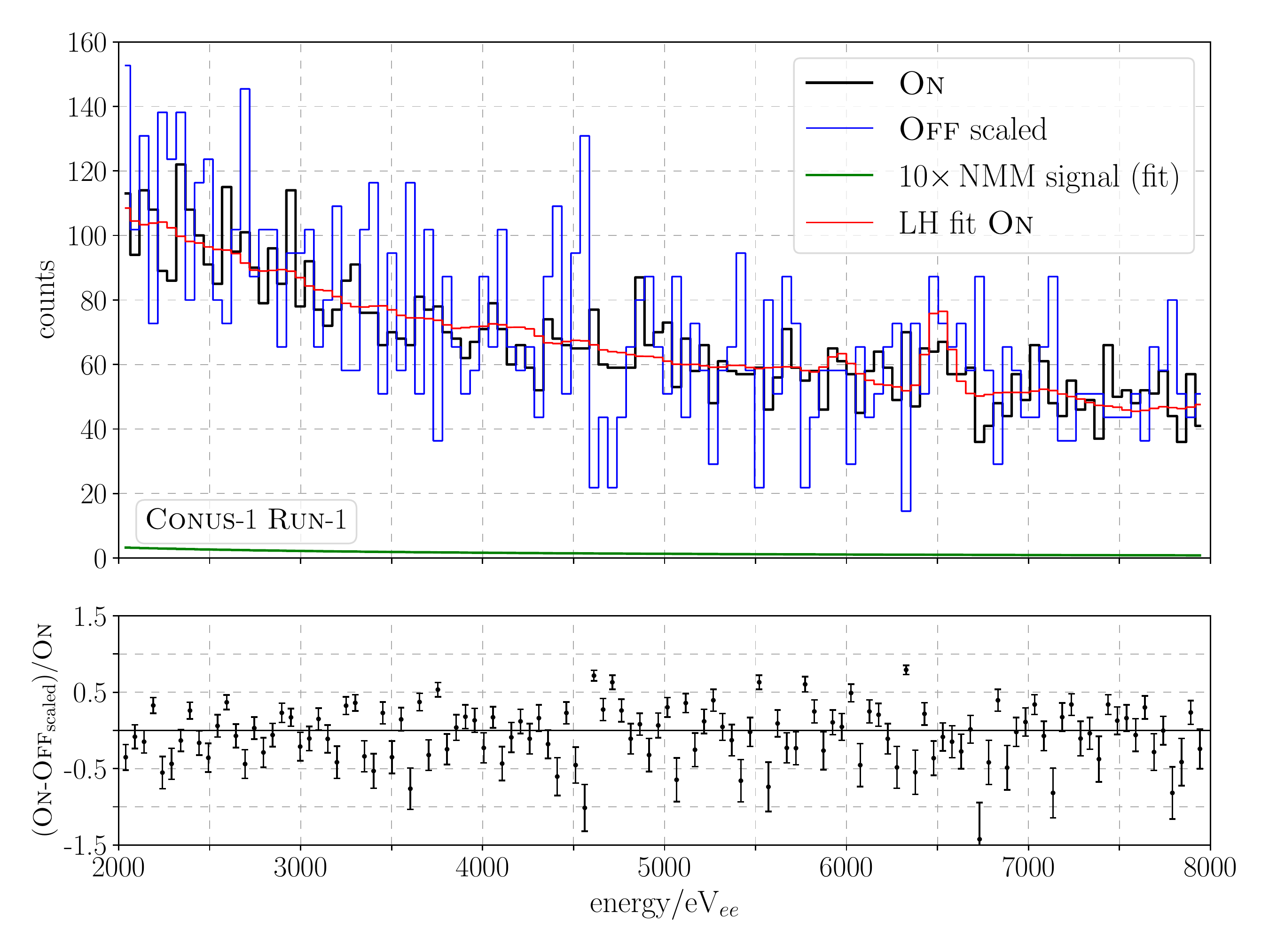}
    \caption{Reactor ON (black) and OFF (blue) data spectra of C1 RUN-1. The OFF data set is scaled to the ON exposure. In the lower part of the figure, the residuals of the data comparison are displayed. No enhancement within the statistics is observed. The red spectrum corresponds to the likelihood fit including the background model and the NMM signal expectation. In green, the NMM signal (multiplied by a factor of 10) from the fit is illustrated.  }
   \label{fig:spectrum_on_vs_off_c1run1}
\end{figure}

No significant enhancement in the reactor ON data over the OFF data is observed within the examined data sets. This is illustrated for C1 in Figure \ref{fig:spectrum_on_vs_off_c1run1}. The reactor ON data is depicted in the ROI together with the scaled reactor OFF data and the resulting residuals. The likelihood fit of the reactor ON data consisting of the background model scaled by the fit parameters as well as the signal expectation of the NMM of 18\,counts is included in the figure. From the complete data set, the following limit on the NMM is derived:
\begin{equation}
    \mu_{\nu} < 7.5\cdot10^{-11}\,\mu_B \text{ (90\% C.L.).}
\end{equation}

\begin{figure}
    \centering
    \includegraphics[width=0.5\textwidth]{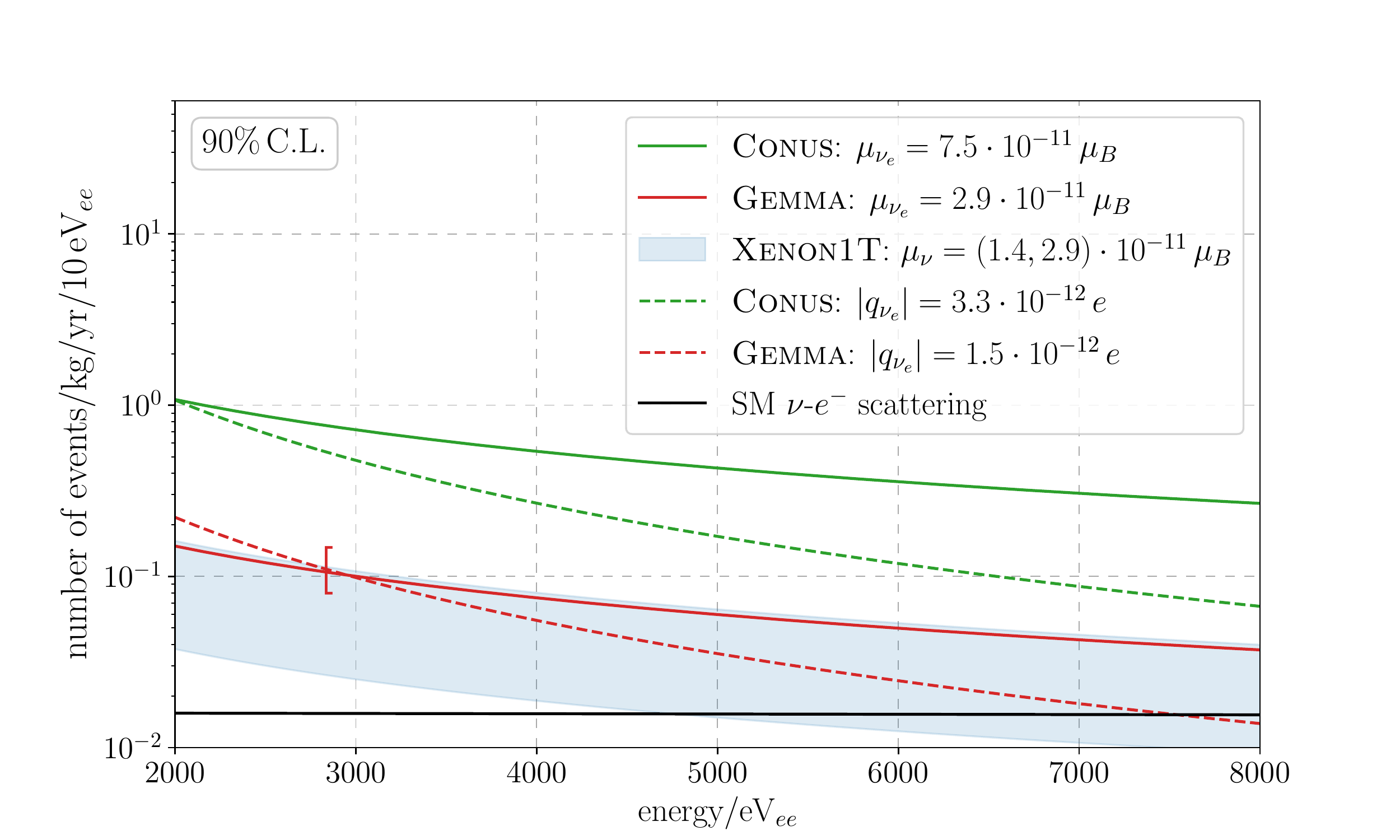}
    \caption{Signal expectations of different NMM values for the data collection period C1R1. 
    We show our current NMM limit in green (corresponding to 299\,counts\,kg$^{-1}$yr$^{-1}$ in the ROI), the current best limit from a reactor site \textsc{(Gemma}: 1133.4 $\mathrm{kg}\cdot\mathrm{d}$ ON, 280.4 $\mathrm{kg}\cdot\mathrm{d}$ OFF, energy threshold of 2.8\,$\mathrm{keV}_{ee}$ indicated by a bracket) in red (corresponding to 45\,counts\,kg$^{-1}$yr$^{-1}$) as well as a potential NMM solution for the detected \textsc{XENON1T} excess with the blue shaded region (corresponding to [10, 45]\,counts\,kg$^{-1}$yr$^{-1}$). The standard model expectation (weak interaction) is indicated in black (9\,counts\,kg$^{-1}$yr$^{-1}$). The bounds on a NMC derived from the upper NMM limits are indicated by dashed lines.
    }
    \label{fig:nmm_result_plot}
\end{figure}

Figure \ref{fig:nmm_result_plot} shows the signal expectation corresponding to the derived bound of 299 counts kg$^{-1}$yr$^{-1}$. It is above the current world's best experimental limit from the \textsc{BOREXINO} experiment \cite{Borexino:2017fbd} and GEMMA experiment \cite{Beda:2013mta} as well as above the potential NMM signal expectation derived from the \textsc{XENON1T} excess \cite{Aprile:2020tmw} by a factor of $\sim$2.6 or more.

The estimated limit on the NMC is derived from the NMM result via Equation~\eqref{eq:millicharge}, while assuming $T\sim2$\,keV corresponding to the lower limit of the ROI. It amounts to 
\begin{equation}
    \vert q_{\nu}\vert < 3.3\cdot10^{-12}\,e_{0} \text{ (90\% C.L.).}
\end{equation}
This limit is about a factor of two above the one obtained from the NMM limit of the reactor antineutrino experiment \textsc{GEMMA}, which was derived in an analogous way~\cite{Studenikin:2013my}. Even though the \textsc{GEMMA} experiment achieved a better NMM limit, the lower minimum recoil energy of 2.0\,keV$_{ee}$, available in the \textsc{CONUS} data analysis, compared to the threshold of 2.8\,keV$_{ee}$ in the \textsc{GEMMA} experiment, leads to a closer approach of the NMC limits.
The signal expectations for both limits are shown in Figure \ref{fig:nmm_result_plot}.


The CONUS constraints on the NMM and the NMC as presented in this publication are limited by the low statistics in particular for the outages, in which the reactor is turned off. Improvements are expected with more data in ON as well as OFF periods. In particular, the experiment will profit from a long background measurement in 2022, since the reactor finished operation by the end of 2021. For all exposure combined, we estimate a sensitivity of $\mu_{\nu}<5.2\cdot10^{-11}\,\mu_B$. Even thought, the exposure will increase significantly, the sensitivity improves by the fourth root leading to an improvement by less than a factor of two. Moreover, it is foreseen to extend the energy range towards higher and lower energy regions. At low energies an improved modeling of background and noise parameters might be achieved at highly stable conditions during recent and upcoming data collection. Optimized DAQ settings will allow to include data at higher energies. Both extensions will further enhance the expected sensitivity. The DAQ system in CONUS was upgraded and improved. Therefore, knowledge on the pulse shape of the signals can be included in future analyses allowing further background understanding and suppression in the ROI. With those significant improvements stronger bounds on electromagnetic properties of neutrinos are expected with the upcoming CONUS data. Furthermore, as soon as a CE$\nu$NS signal is observed, additional limits can also be derived from its nuclear recoil spectrum in the sub-keV regime.

\section{Acknowledgements}
We thank all the technical and administrative staff who helped building the experiment, in particular
the MPIK workshops and Mirion Technologies (Canberra) in Lingolsheim. 
We express our gratitude to the Preussen Elektra GmbH for great support and for 
hosting the \textsc{CONUS} experiment. 
We thank Dr.~S.~Schoppmann (MPIK) for assistance on the analysis and Dr.~M.~Seidl
(Preussen Elektra) for providing simulation data on the fission rate evolution 
over a reactor cycle. 
The \textsc{CONUS} experiment is supported financially by the Max Planck Society (MPG), 
T.~Rink by the German Research Foundation (DFG) through the research training 
group GRK 1940, and together with J.~Hakenm\"uller by the IMPRS-PTFS.


\bibliographystyle{spphysmod} 
\bibliography{literature.bib}

\providecommand{\noopsort}[1]{}\providecommand{\singleletter}[1]{#1}%
\begin{thebibliography}{10}
\providecommand{\url}[1]{{#1}}
\providecommand{\urlprefix}{URL }
\expandafter\ifx\csname urlstyle\endcsname\relax
  \providecommand{\doi}[1]{DOI \discretionary{}{}{}#1}\else
  \providecommand{\doi}{DOI \discretionary{}{}{}\begingroup
  \urlstyle{rm}\Url}\fi

\bibitem{PhysRevLett.95.151802}
N.F. Bell, et~al., Phys. Rev. Lett. \textbf{95}, 151802 (2005).
\newblock \doi{10.1103/PhysRevLett.95.151802}

\bibitem{Fujikawa:1980yx}
K.~Fujikawa, R.~Shrock, Phys. Rev. Lett. \textbf{45}, 963 (1980).
\newblock \doi{10.1103/PhysRevLett.45.963}

\bibitem{PhysRevD.39.3378}
P.~Vogel, J.~Engel, Phys. Rev. D \textbf{39}, 3378 (1989).
\newblock \doi{10.1103/PhysRevD.39.3378}

\bibitem{Giunti:2014ixa}
C.~Giunti, A.~Studenikin, Rev. Mod. Phys. \textbf{87}, 531 (2015).
\newblock \doi{10.1103/RevModPhys.87.531}

\bibitem{shrock1982electromagnetic}
R.E. Shrock, Nuclear Physics B \textbf{206}(3), 359 (1982)

\bibitem{Bell:2006wi}
N.F. Bell, et~al., Phys. Lett. B \textbf{642}, 377 (2006).
\newblock \doi{10.1016/j.physletb.2006.09.055}

\bibitem{cisneros1971effect}
A.~Cisneros, Astrophysics and Space Science \textbf{10}(1), 87 (1971)

\bibitem{barbieri1988limit}
R.~Barbieri, R.N. Mohapatra, Physical Review Letters \textbf{61}(1), 27 (1988)

\bibitem{Borexino:2017fbd}
M.~Agostini, et~al., Phys. Rev. D \textbf{96}(9), 091103 (2017).
\newblock \doi{10.1103/PhysRevD.96.091103}

\bibitem{Beda:2013mta}
A.G. Beda, et~al., Phys. Part. Nucl. Lett. \textbf{10}, 139 (2013).
\newblock \doi{10.1134/S1547477113020027}

\bibitem{TEXONO:2006xds}
H.T. Wong, et~al., Phys. Rev. D \textbf{75}, 012001 (2007).
\newblock \doi{10.1103/PhysRevD.75.012001}

\bibitem{Derbin:1993wy}
A.I. Derbin, et~al., JETP Lett. \textbf{57}, 768 (1993)

\bibitem{COHERENT:2017ipa}
D.~Akimov, et~al., Science \textbf{357}(6356), 1123 (2017).
\newblock \doi{10.1126/science.aao0990}

\bibitem{Miranda:2019wdy}
O.G. Miranda, et~al., JHEP \textbf{07}, 103 (2019).
\newblock \doi{10.1007/JHEP07(2019)103}

\bibitem{billard2018prospects}
J.~Billard, et~al., Journal of Cosmology and Astroparticle Physics
  \textbf{2018}(11), 016 (2018)

\bibitem{Allen:1992qe}
R.C. Allen, et~al., Phys. Rev. D \textbf{47}, 11 (1993).
\newblock \doi{10.1103/PhysRevD.47.11}

\bibitem{Auerbach:2001wg}
L.B. Auerbach, et~al., Phys. Rev. D \textbf{63}, 112001 (2001).
\newblock \doi{10.1103/PhysRevD.63.112001}

\bibitem{Canas:2015yoa}
B.C. Canas, et~al., Phys. Lett. B \textbf{753}, 191 (2016).
\newblock \doi{10.1016/j.physletb.2015.12.011}.
\newblock [Addendum: Phys.Lett.B 757, 568--568 (2016)]

\bibitem{Aprile:2020tmw}
E.~Aprile, et~al., Phys. Rev. D \textbf{102}(7), 072004 (2020).
\newblock \doi{10.1103/PhysRevD.102.072004}

\bibitem{Arceo-Diaz:2015pva}
S.~{Arceo-D{\'i}az}, et~al., Astropart.Phys. \textbf{70}, 1 (2015).
\newblock \doi{10.1016/j.astropartphys.2015.03.006}

\bibitem{Viaux:2013lha}
N.~Viaux, et~al., Phys. Rev. Lett. \textbf{111} (2013).
\newblock \doi{10.1103/PhysRevLett.111.231301}

\bibitem{Kuznetsov:2009zm}
A.V. Kuznetsov, et~al., Int.J.Mod.Phys.A \textbf{24}, 5977 (2009).
\newblock \doi{10.1142/S0217751X09047612}

\bibitem{giunti:2007}
C.~Giunti, C.W. Kim, \emph{Fundamentals of Neutrino Physics and Astrophysics,
  Chapter 3}, 1st edn. ({Oxford Univ. Press}, 2007)

\bibitem{Studenikin:2013my}
A.~Studenikin, EPL \textbf{107}(2), 21001 (2014).
\newblock \doi{10.1209/0295-5075/107/21001}.
\newblock [Erratum: EPL 107, 39901 (2014)]

\bibitem{Bonet:2020awv}
H.~Bonet, et~al., Phys. Rev. Lett. \textbf{126}(4), 041804 (2021).
\newblock \doi{10.1103/PhysRevLett.126.041804}

\bibitem{Bonet:2020ntx}
H.~Bonet, et~al., Eur. Phys. J. C \textbf{81}(3), 267 (2021).
\newblock \doi{10.1140/epjc/s10052-021-09038-3}

\bibitem{bonet2021bkg}
H.~Bonet, et~al., arXiv:2112.09585  (2021)

\bibitem{hakenmuller2019neutron}
J.~Hakenm{\"u}ller, et~al., Eur. Phys. J. C \textbf{79}(8), 1 (2019)

\bibitem{beda2007first}
A.G. Beda, et~al., Physics of Atomic Nuclei \textbf{70}(11), 1873 (2007)

\bibitem{Huber:2011wv}
P.~Huber, Phys. Rev. C \textbf{84}, 024617 (2011).
\newblock \doi{10.1103/PhysRevC.85.029901}.
\newblock [Erratum: Phys. Rev. C 85, 029901 (2012)]

\bibitem{Mueller:2011nm}
T.A. M{\"u}ller, et~al., Phys. Rev. C \textbf{83}, 054615 (2011).
\newblock \doi{10.1103/PhysRevC.83.054615}

\bibitem{Kopeikin:2003gu}
V.~Kopeikin, et~al., Phys. Atom. Nucl. \textbf{67}, 1963 (2004).
\newblock \doi{10.1134/1.1825513}

\bibitem{DayaBay:2016ssb}
F.P. An, et~al., Chin. Phys. C \textbf{41}(1), 013002 (2017).
\newblock \doi{10.1088/1674-1137/41/1/013002}

\bibitem{perkins1991tables}
S.~Perkins, et~al., \emph{Tables and graphs of atomic subshell and relaxation
  data derived from the LLNL Evaluated Atomic Data Library (EADL), Z= 1--100}
  (US Department of Energy, Office of Scientific and Technical Information,
  1991)

\bibitem{bonet2021novel}
H.~Bonet, et~al., JHEP \textbf{05}, 085 (2022).
\newblock \doi{10.1007/JHEP05(2022)085}

\end{thebibliography}
\end{document}